\documentclass[twocolumn,preprintnumbers]{revtex4}%
\usepackage{amsmath}
\usepackage{amssymb}
\usepackage{graphicx}
\usepackage{dcolumn}
\usepackage{bm}
\usepackage{amsfonts}%
\begin{document}
\title{Comment on "Topological Transitions in Berry's Phase Interference Effects"}
\author{Eyal Buks}
\affiliation{Department of Electrical Engineering, Technion, Haifa 32000, Israel }
\date{\today }
\maketitle





The paper by Lyanda-Geller (henceforth LG) \cite{LG} predicts a variation from
$\pi$ to zero of Berry's phase, which may manifest itself in a step-like
current-magnetic field and current-gate voltage characteristics predicted for
in-plane magnetoresistance of rings in noncentrosymmetric materials.

As a demonstrating example LG considers a spin $1/2$ evolving according to the
following Schr\"{o}dinger equation%

\begin{equation}
\frac{d}{dt}\left\vert \psi\right\rangle =i\mathbf{\Omega}\cdot\mathbf{\sigma
}\left\vert \psi\right\rangle , \label{Schro}%
\end{equation}
where $\mathbf{\sigma}$ is Pauli matrix vector, and the vector $\mathbf{\Omega
}$, which lies in the $xy$ plane, is given by%

\begin{equation}
\mathbf{\Omega}=\left(  \omega_{1}-\omega_{0}\cos\left(  \omega t\right)
,\omega_{0}\sin\left(  \omega t\right)  ,0\right)  ,
\end{equation}
where $\omega$, $\omega_{0}$, $\omega_{1}$ are constants.

The state at the initial time $t_{0}=-\pi/\omega$ is assumed to be an
eigenstate of the instantaneous Hamiltonian $\mathcal{H}\left(  t_{0}\right)
$. \ In general, if $\omega_{1}\neq\omega_{0}$ and $\omega$ is sufficiently
small the system is expected to evolve adiabatically, namely, at any later
time $t>t_{0}$ the state $\left\vert \psi\left(  t\right)  \right\rangle $
will remain approximately an eigenvector of the instantaneous Hamiltonian
$\mathcal{H}\left(  t\right)  $. \ However, LG further claims that adiabatic
evolution is possible also for the case $\omega_{1}=\omega_{0}$ provided that
$\omega$ is sufficiently small. \ Consequently, LG concludes that the
conductance may exhibit an abrupt jump as the parameter $\omega_{1}$ is varied
across the point $\omega_{1}=\omega_{0}$.

In this comment we claim that, contrary to Ref. \cite{LG}, the conductance
steps predicted by LG are not abrupt but rather they occur along a finite
range. \ In general, such abrupt jumps are ruled out since the system under
consideration has a linear respond \cite{Buks Upper}. \ For a finite time
interval, the change in the final state of the system cannot remain finite in
the limit where the perturbation causing the change (modifying $\omega_{1}$)
approaches zero. \ This implies that the change in conductance occurs along a
finite range.

To probability $P_{+-}^{\left(  t_{1}\right)  }$ in Eq. (8) of Ref. \cite{LG}
is calculated correctly. \ Indeed, for the case $\omega_{1}=\omega_{0}$, the
state of the system will remain nearly unchanged as the curve $\mathbf{\Omega
}\left(  t\right)  $ crosses the degeneracy point at the origin provided that
$\omega$ is sufficiently small. \ However, across this point the local
eigenvectors $\left\vert n_{\pm}\right\rangle $ change abruptly, namely,
$\left\vert n_{+}\right\rangle $ becomes $\left\vert n_{-}\right\rangle $ and
vise versa. \ Therefore, $P_{+-}^{\left(  t_{1}\right)  }$ in Ref. \cite{LG}
is not the probability to have a state mixing (Zener transition), but rather
the probability \textit{not} to have one.

\begin{figure}
[ptb]
\begin{center}
\includegraphics[
height=5.713in,
width=2.2727in
]%
{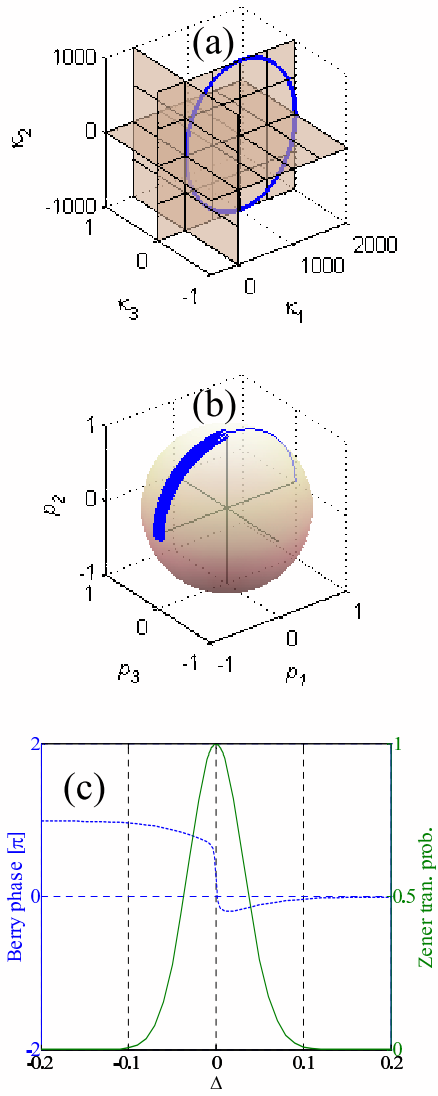}%
\caption{Numerical integration of Eq. \ref{Schro}.}%
\label{Nu int}%
\end{center}
\end{figure}

To further support this conclusion we integrate Eq. \ref{Schro} numerically
from $t_{0}=-\pi/\omega$ to $t_{1}=\pi/\omega$ \cite{Buks Modu}. \ For the
example depicted in Fig. \ref{Nu int} below we chose $\omega_{0}/\omega=1000$.
\ In Fig. \ref{Nu int} (a) the vector $\mathbf{\Omega}\left(  t\right)  $ is
depicted for the case $\Delta\equiv\left(  \omega_{1}-\omega_{0}\right)
/\omega_{0}=0.0015$, and in Fig. \ref{Nu int} (b) the polarization vector
$\left\langle \psi\right\vert \mathbf{\sigma}\left\vert \psi\right\rangle $ is
seen for the same value of $\Delta$. \ For this example the state at the final
time $t_{1}$ is nearly orthogonal to the initial state at time $t_{0}$,
indicating that a state mixing (Zener transition) occurs. \ However for larger
values of $\left\vert \Delta\right\vert $, adiabaticity is restored, as can be
seen in Fig. \ref{Nu int} (c), where the numerically calculated probability of
Zener transition is plotted as a function of $\Delta$. \ The same plot also
shows the Berry's phase $\gamma_{B}$ plotted vs. $\Delta$. \ Near $\Delta=0$
indeed $\gamma_{B}$ changes by $\pi$ as predicted by LG, however, this occurs
along a finite range of $\Delta$ rather than abruptly.%

\newpage
\bibliographystyle{plain}
\bibliography{apssamp}

\end{document}